\documentclass[12pt,preprint]{emulateapj}

\shorttitle{Growing transverse oscillations of a multistranded loop}
\shortauthors{Wang et al.}

\begin{document}

\title{Growing transverse oscillations of a multistranded loop
  observed by SDO/AIA}

\author{Tongjiang Wang\altaffilmark{1,2}, Leon Ofman\altaffilmark{1,2,3}, 
Joseph M. Davila\altaffilmark{2}, and Yang Su\altaffilmark{1,2,4}}
\altaffiltext{1}{Department of Physics, Catholic University of America,
   620 Michigan Avenue NE, Washington, DC 20064, USA; tongjiang.wang@nasa.gov}
\altaffiltext{2}{NASA Goddard Space Flight Center, Code 671, Greenbelt, MD 20770, USA}
\altaffiltext{3}{Visiting Associate Professor, Tel Aviv University, Israel}
\altaffiltext{4}{IGAM/Department of Physics, University of Graz, Universitaetsplatz 5, 
Graz 8010, Austria}

\begin{abstract}
The first evidence of transverse oscillations of a multistranded loop with growing amplitudes
and internal coupling observed by the Atomspheric Imaging Assembly (AIA) onboard the Solar 
Dynamics Observatory (SDO) is presented. The loop oscillation event occurred on 2011 March 8, 
triggered by a CME. The multiwavelength analysis reveals the presence of multithermal strands
in the oscillating loop, whose dynamic behaviors are temperature-dependent, 
showing differences  in their oscillation amplitudes, phases and emission evolution. 
The physical parameters of growing oscillations of two strands in 171 \AA\ are measured and 
the 3-D loop geometry is determined using STEREO-A/EUVI data. These strands have very similar
frequencies, and between two 193 \AA\ strands a quarter-period phase delay sets up. These
features suggest the coupling between kink oscillations of neighboring strands and the
interpretation by the collective kink mode as predicted by some models. However, the temperature 
dependence of the multistarnded loop oscillations was not studied previously and needs further
investigation. The transverse loop 
oscillations are associated with intensity and loop width variations. We suggest that 
the amplitude-growing kink oscillations may be a result of continuous 
non-periodic driving by magnetic deformation of the CME, which deposits energy into the loop 
system at a rate faster than its loss. 
\end{abstract}

\keywords{Sun: Flares --- Sun: corona --- Sun: oscillations --- waves --- 
Sun: UV radiation }

\section{Introduction}
 Transverse coronal loop oscillations have been extensively studied in both observation
and theory \citep[see recent reviews by][]{asc09, ter09, rud09}.
Observations from TRACE and STEREO/EUVI show that these oscillations are triggered by 
a flare or a coronal mass ejection (CME) \citep[e.g.,][]{asc02, asc09}. 
They have been interpreted as fast standing magnetohydrodynamic (MHD) kink modes \citep{asc99, nak99}. 
Transverse loop oscillations are often observed with a rapid decay within
several periods \citep{nak99, asc02, whi12}. Sometimes the undamped 
oscillations are observed \citep{asc02, asc11}. It has been suggested that the expected damping is 
balanced by amplification due to cooling \citep{rud11a,rud11b, rud11c}. Moreover, 
transverse oscillations are observed not only in single loops but also in a bundle of loops 
\citep{ver04, ofm08, ver09, asc11}. Recent theories
have shown that the global kink mode still exists in models with multiple strands,
but its transverse dynamics are influenced by the internal fine structure due to the 
coupling and phase mixing of neighboring strands in properties such as the frequency and damping 
\citep{ofm05, ofm09, lun08, lun09, lun10, ter08, van08}. These studies have significantly contributed
to the progress of coronal seismology, a diagnostic tool to probe the physical parameters in
the corona \citep[e.g.,][]{nak05}. 

Here we present the first example of transverse oscillations of a multistranded loop 
observed by the Atmospheric Imaging Assembly (AIA) on the Solar Dynamics Observatory (SDO), 
showing the evidence for growing amplitudes and the internal coupling.

\section{Observations}
The oscillation event occurred on 2011 March 8, 19:40$-$20:40 UT in AR 11165 on 
the limb, observed with SDO/AIA. An M1.5 GOES-class flare associated with a CME 
and a surge were also observed during this time. The AIA records continuous images of the 
full Sun with 1.5$^{''}$ resolution and 12 s cadence \citep{lem11}. This flare-CME event was first
studied by \citet{su12} using AIA, STEREO-A and RHESSI data. We present the analysis
of loop oscillations using data from four AIA bands, 171, 193, 211, and 304 \AA, 
as well as STEREO-A/EUVI data.

\begin{figure*}
\epsscale{1.0}
\plotone{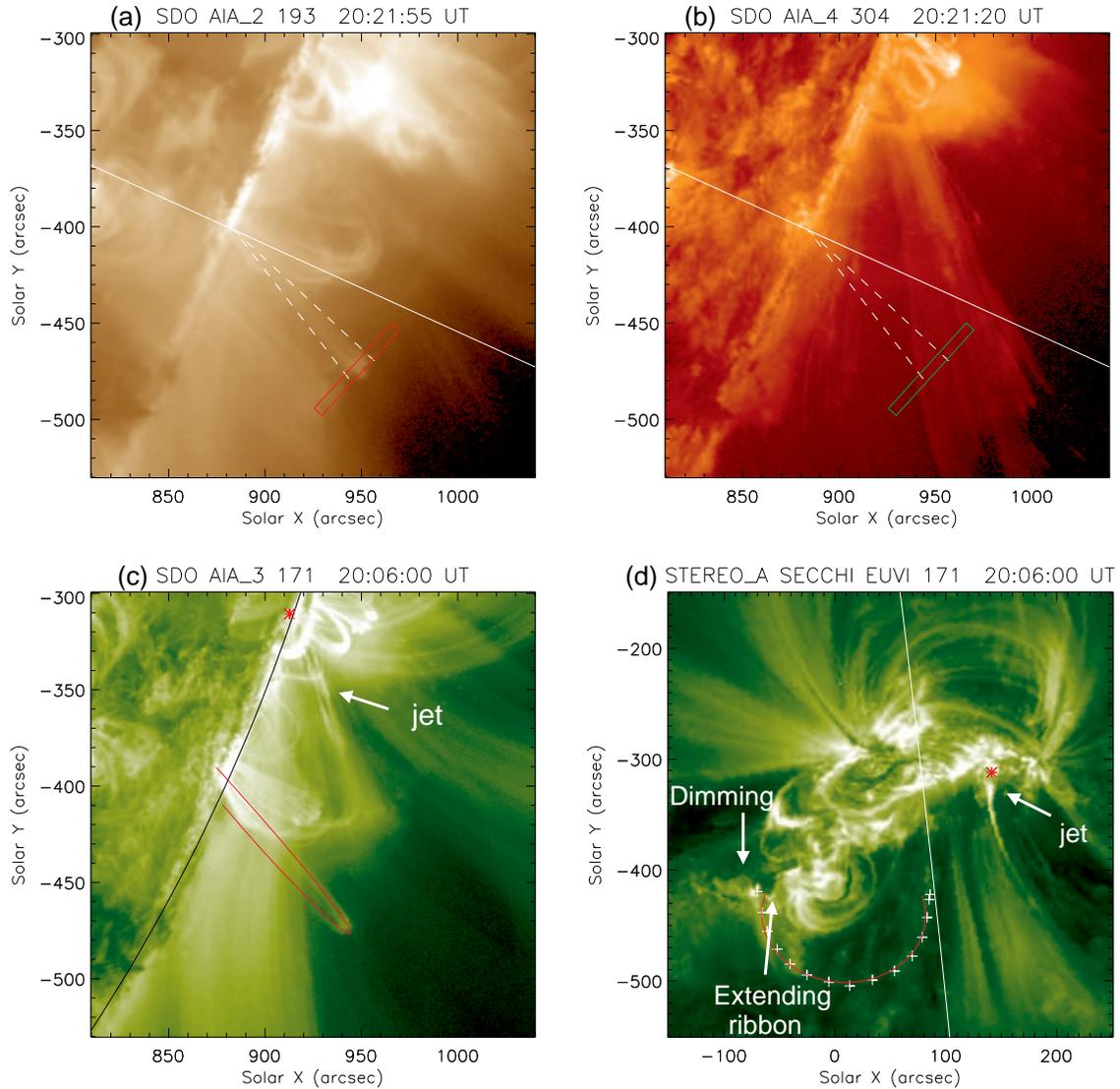}
\caption{ \label{fgmap} Observations of the transverse loop oscillation event
on 2011 March 08, with SDO/AIA in (a) 193 \AA\ and (b) 304 \AA\ bands. A small narrow box shows a cut 
used for stack plots. The dashed lines outline the oscillating loop seen in 193 \AA. The solid line shows 
the solar radial direction passing the loop's footpoint. (c) and (d): Simultaneous observations 
of the oscillating loop with AIA 171 \AA\ and STEREO-A/EUVI 171 \AA\ at 20:06:00 UT. In (d), the red curve 
is the best fit of a circular loop model to the oscillating loop (outlined with {\it pluses}). 
The white line shows the limb as seen from SDO with the visible disk on its left. The symbol of 
{\it asterisk} marks the footpoint of the jet. In (c) the lines and the symbol have the same
meanings as in (d) but for the SDO view. [See animations in the online journal]}
\end{figure*}

\section{Results}
\begin{figure}
\epsscale{1.0}
\plotone{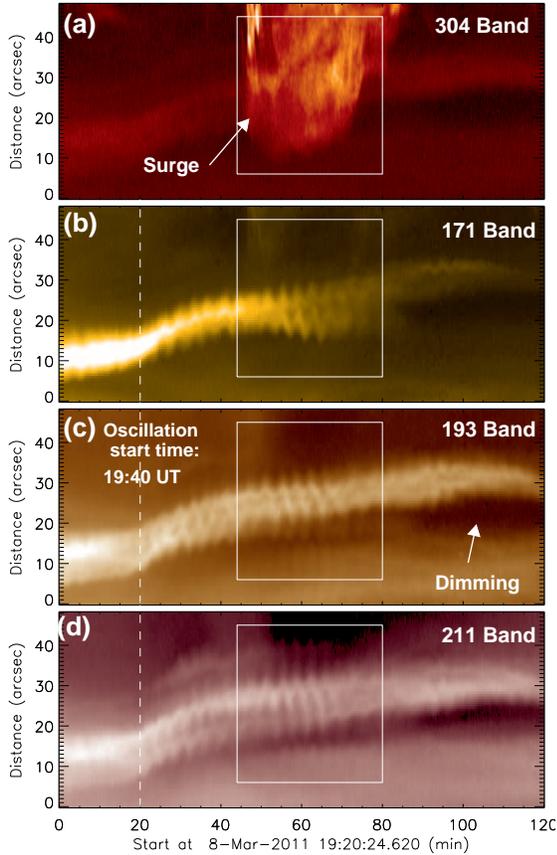}
\caption{ \label{fgcut} Time-distance maps along a cut at the loop apex 
as shown in Figure~\ref{fgmap}(a) (averaged over the narrow width) 
in four bands, (a) 304 \AA, (b) 171 \AA, (c) 193 \AA, and (d) 211 \AA. A box marks the time 
period of interest. The vertical dashed lines indicate the start time of transverse
loop oscillations.}
\end{figure}

The oscillating loop is visible in AIA 171, 193 and 211 \AA\ bands (Figure~\ref{fgmap}). 
The loop plane is almost parallel to the line-of-sight. In the 304 \AA, a surge was observed
passing by the loop, and could have triggered its oscillations (Figure~\ref{fgmap}(b)). 
The surge began at 19:55 UT with the velocity of 170$-$230 km~s$^{-1}$, and the ejected 
material fell back on the Sun at 20:30 UT at about 80 km~s$^{-1}$. However, the STEREO-A 
observations viewing the AR at the disk center show that this surge was not directly related
to the loop oscillation. Figure~\ref{fgcut} shows the flux
evolution of the four bands along a cut at the loop's apex. The loop 
started to oscillate at about 19:40 UT with a rapid drift towards the
north (flare source). The start time of oscillations is consistent with the CME 
acceleration time \citep[see][]{su12}. The oscillations became evident during the period 
of the surge. In this period, the loop in 171 \AA\ gradually split
into two strands manifesting the unusual oscillations with growing amplitudes. 
The loop in 193 and 211 \AA, composed of several close strands, shows the oscillations
with no clear change in amplitude. The difference of these strands in spatial distribution 
and temporal evolution indicates that the loop consists of the multithermal structure. 
The lower 171 \AA\ strand disappeared after the surge, while a dimming formed in 193 and 211 \AA. 
This may suggest that parts of the loop erupted. The upper branch of the loop dimmed slowly 
in 171 \AA, while remained bright in 193 and 211 \AA, suggesting possible heating.

\begin{figure*}
\epsscale{1.0}
\plotone{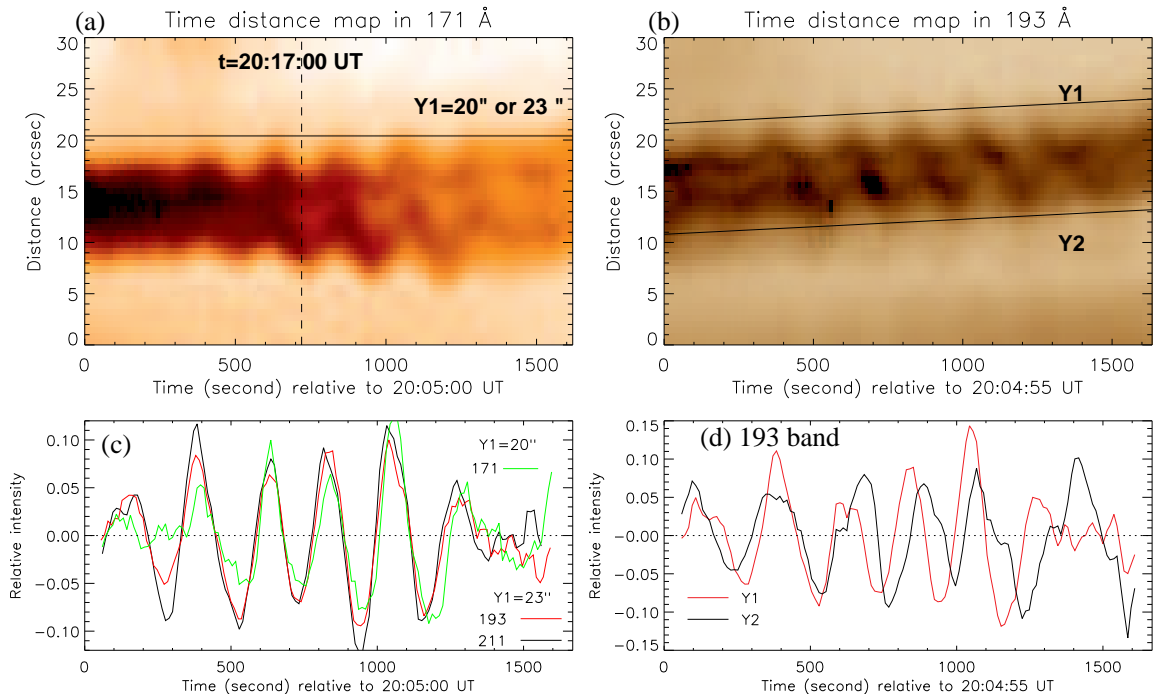}
\caption{ \label{fgosc} Time-slice diagrams of (a) 171 \AA\  and (b) 193 \AA\ flux of the 
oscillating loop (in negative color). The vertical dashed line marks the time 
for Figure~\ref{fgpar}(b). (c) Time profiles of the 171, 193
and 211 \AA\ relative intensities at position Y1 (marked in (a)). (d) Time profiles of the 193 \AA\ 
relative intensities at two positions Y1 and Y2 (marked in (b)), where for a clear comparison
the light curve for Y2 (solid black line) is plotted as its negative. In (c) and (d), the 
relative intensities (with the background subtracted) are normalized to 
the smoothed background trend. }
\end{figure*}

\begin{figure*}
\epsscale{1.}
\plotone{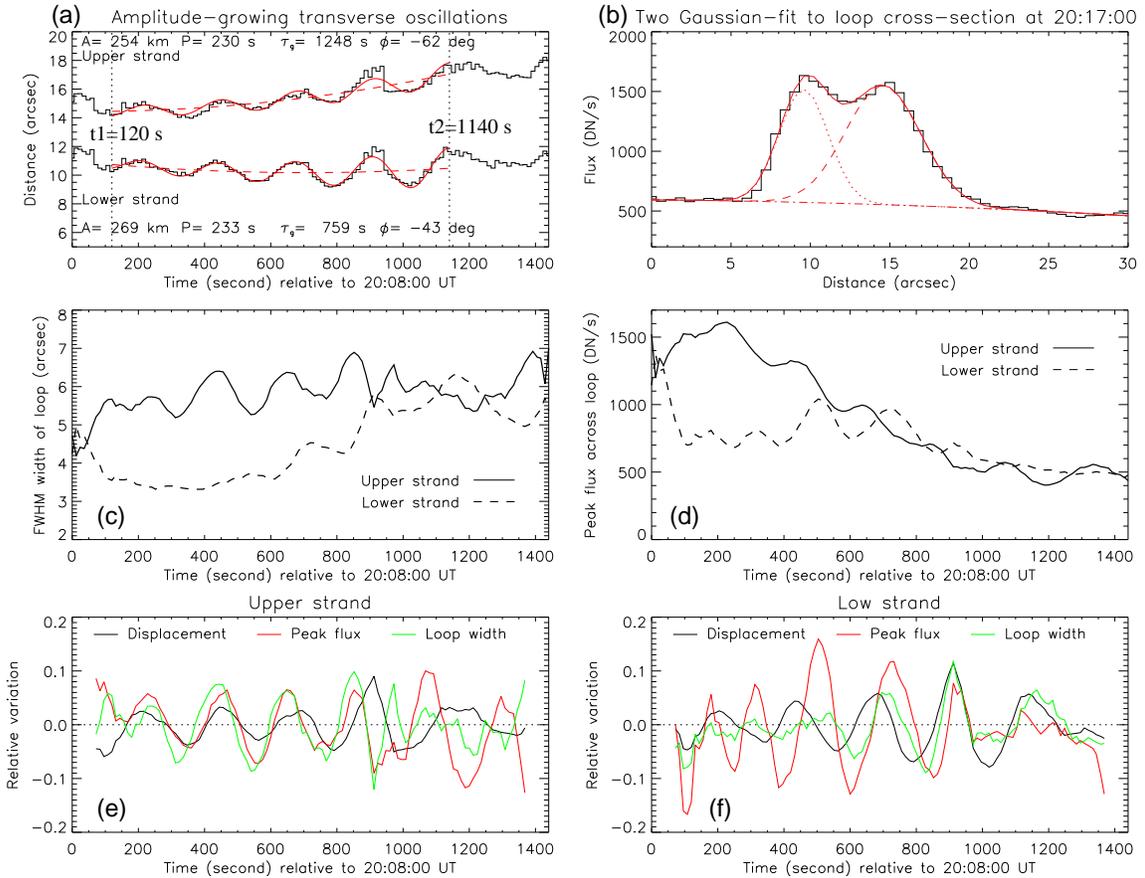}
\caption{ \label{fgpar} (a) The displacement oscillations of two loop strands in 171 \AA,
and the best fits (red lines) with Equation~(\ref{eqosc}). The red dashed lines are the parabolic fit 
to a drift. (b) Loop cross-sectional flux profile at 20:17:00 UT, and the double Gaussian fit
(red solid line), where the dotted and dashed lines show the fitted components for 
the lower and upper strands, and the dot-dashed line for the fitted background.
(c) Loop FWHM width, and (d) cross-sectional peak flux of the upper (solid line) and lower
(dashed line) strands as a function of time. (e) and (f): The normalized relative variations 
of displacement (black line), cross-sectional peak flux (red line), and loop FWHM width 
(green line) for the upper and lower strands, respectively. }
\end{figure*}

We examine the phase relation between the oscillations of different strands. 
Figure~\ref{fgosc}(a) shows the oscillations of two 171 \AA\ strands almost in phase 
(see also Figure~\ref{fgpar}(a)). Figure~\ref{fgosc}(b) shows the oscillations of 
two 193 \AA\ strands initially in phase, but becoming shifted by a quarter-period 
after two periods. This behavior can be clearly seen from a comparison
of the intensity evolution at two locations (with a linear drift of 1 km~s$^{-1}$) 
near the displacement maxima of the strands (Figure~\ref{fgosc}(d)). 
For the upper strand, the oscillation period is estimated
to be 258$\pm$66 s for the initial phase, and 216$\pm$15 s for the later phase,
while for the lower strand, the period is 294$\pm$22 s for the initial phase, 
and 216$\pm$83 s for the later phase, where the periods are measured from the average of
time intervals of the intensity peaks and the errors the standard deviation. 
This evolution suggests the setup of
collective oscillation between the two neighboring strands, which has a slightly
shorter period than before the coupling. Figure~\ref{fgosc}(c) compares the
flux evolution in three bands at the location (Y1) near the upper strands, where 
Y1=20$^{''}$ in 171 \AA\, and Y1=23$^{''}$ in 193 and 211 \AA\ considering their slight
offsets in position. It shows that the upper strands seen in the three
bands oscillated with similar periods and nearly in phase. 
The oscillations of these strands with certain phase shifts and the similar 
frequencies suggest collective dynamics of a multistranded 
loop system \citep{lun08, lun09}.

We measure the time variation of oscillation amplitudes for two 171 \AA\ strands by locating
the cross-sectional flux maxima using a double Gaussian best fit with a parabolic background. 
Figure~\ref{fgpar}(b) shows an example for the fitting of emission profile across the loop. 
The measured loop displacements, FWHM width, and cross-sectional peak flux as a function of 
time are shown in Figures~\ref{fgpar} (a), (c), and (d), respectively.
The upper strand has the average FWHM diameter of 4.2$\pm$0.4 Mm, and the lower strand has
that of 3.3$\pm$0.7 Mm. By fitting the displacement oscillations with an amplitude-growing 
sine function with a parabolic drift given by,
\begin{equation}
 a(t)=A\,{\rm sin}(\frac{2\pi{(t-t_0)}}{P}+\phi){\rm e}^{\frac{t-t_0}{\tau_g}}+a_0+a_1(t-t_0)+a_2(t-t_0)^2,
 \label{eqosc}
\end{equation} 
we determine the parameters of the oscillation: 
amplitude ($A$), period ($P$), phase ($\phi$), and amplitude growth time 
scale ($\tau_g$), where $t_0$ is the start time of analyzed oscillations. The measured parameters 
are shown on the plots and listed in Table~\ref{tabpar}. The growing oscillations 
of the two strands started and ended almost simultaneously, lasting over four periods. 
The lower strand has higher growth rate of the oscillation amplitude than the upper one. We estimate
the increase in their amplitudes by $A_2/A_1$=3.8 for the lower strand, and $A_2/A_1$=2.3 for 
the upper strand using $A_2/A_1$=$e^{\Delta{t}/\tau_g}$ with the life time $\Delta{t}$=1020 s.

\begin{deluxetable}{lccccccc}
 \tabletypesize{\scriptsize}
 \tablecaption{Physical parameters of the amplitude-growing oscillations in the 171 \AA\ band\tablenotemark{a} \label{tabpar}}
 \tablewidth{0pt}
 \tablehead{
 \colhead{Loop} & \colhead{$P$} & \colhead{$A$} & \colhead{$\tau_g$} & \colhead{$\phi$} & \colhead{$A_2/A_1$} & \colhead{$d$} & \colhead{$L$}\\
 \colhead{} & \colhead{(s)} & \colhead{(Mm)} & \colhead{(s)} & \colhead{($^{\circ}$)} &  & \colhead{(Mm)} & \colhead{(Mm)}}
\startdata
upper & 230 & 0.254 & 1248 & $-$62 & 2.3 & 4.2$\pm$0.4 & 77\\
lower & 233 & 0.269 & 759  & $-$43 & 3.8 & 3.3$\pm$0.7 & 77\\
\enddata
\tablenotetext{a}{$P$--oscillation period, $A$--amplitude at the start time, $\tau_g$--amplitude growth time,
$\phi$-phase, $A_2/A_1$--ratio of amplitudes at the end and start time of analyzed oscillations, 
$d$--loop FWHM diameter, and $L$--loop length.}
\end{deluxetable}

The displacement oscillations are found in association with intensity and loop
width fluctuations. Figures~\ref{fgpar}(e) and (f) show comparisons between the relative 
displacement, cross-sectional peak flux, and loop width time variations, 
where a 290~s smoothed trend for all parameters has been subtracted, the relative displacements
are normalized to a scale of 7 Mm, and the relative peak flux and loop width are normalized to
their smoothed trend. An inphase relationship is found between the loop width and intensity 
fluctuations (with relative amplitudes of $\sim$5\%$-$15\%) for both strands.
The phase relationship between displacement and intensity oscillations is different 
for the two strands, being in-phase for the upper strand and a quarter-period shift 
for the lower strand. 

To determine the trigger of the oscillation and measure the loop geometry, 
we analyze the STEREO-A observations (Figure~\ref{fgmap}(d)). A jet which corresponds to
the surge in AIA 304 \AA\ band can be easily identified, whereas the direct identification 
of the oscillating loop in STEREO/EUVI images is not obvious due to the low (75 s) cadence. 
We use the following procedure to locate the oscillating loop. 
First we speculate that a faint loop (outlined in Figure~\ref{fgmap}(d)) is the target
based on its evolutionary features. The movies (available in online version of this letter) 
show that this loop was apparently shrinking in STEREO/EUVI, suggesting a correspondence to the change 
in inclination of the oscillating loop in SDO/AIA. In addition, this STEREO loop dimmed 
simultaneously as the AIA loop (at about 20:40 UT). Next we model this STEREO loop with 
a 3D arc and map it onto the AIA view to compare with the observation. The method is similar 
to that used by \citet{asc02}. A circular loop model is made by optimizing two free parameters, 
$h_0$ and $\theta$, where $h_0$ is the height of the circular loop 
center above the solar surface and $\theta$ is the inclination angle of the loop plane to
vertical. The fact that the best fit loop model, when mapped to the AIA view, matches
the observation confirms the initial conjecture. The calculated loop parameters 
are $h_0$=18 Mm, $\theta$=24$^{\circ}$, the curvature radius $r$=59 Mm, and the loop length
$L$=212 Mm. Taking the period $P$=230 s, we estimate the phase speed of the oscillations 
in the fundamental mode, $V_p$=2L/P=1840 km~s$^{-1}$. For the kink mode \citep{rob84}, we obtain 
the Alfv\'{e}n speed $V_A$=1360 km~s$^{-1}$ if assuming the phase speed ($V_p$) equal to 
the kink speed ($C_k$) and the loop density contrast of 10,  and estimate the average magnetic field 
in the loop, $B$=6$-$20 G, for typical coronal loop densities ($10^8-10^9$ cm$^{-3}$). 

The STEREO-A observations suggest that the loop oscillation was triggered by the CME but
not the surge/jet. The EUVI 171, 195 and 284 \AA\ bands observed the eruption of a large flux 
rope at 19:40 UT, which appeared as a CME at 21:12 UT in SOHO/LASCO C2. In 171 \AA\ 
a bright ribbon appeared at about 20:00 UT, and extended towards the left footpoint of 
the oscillating loop, followed by the formation of a dimming region 
(Figure~\ref{fgmap}(d), and movies in online version). 
The dimmings observed in the EUV and/or soft X-ray range were interpreted as the coronal 
plasma evacuation at the footpoints of a magnetic flux rope when it rapidly opens or expands 
\citep[e.g.,][]{ste97, wan02}. The presence of a bright ribbon at the boundary 
of the extending dimming region has not been reported previously in literature. We suggest that
it could be caused by interaction (via local reconnection) between the expanding magnetic
fields of the CME and the ambient closed magnetic loops. \citet{ima07} found 
the temperature-dependent strong upflows (up to $\sim$150 km~s$^{-1}$) at
the boundary of the dimming region with Hinode/EIS, supporting this suggestion. 
Therefore, continuous magnetic interaction by the CME may drive 
the loop oscillation with growing amplitudes, and also lead to the heating 
(e.g. by hot outflows) and partial eruption of the oscillating loop, as observed. 
This scenario is also supported by a coincidence of the flux rope eruption with the excitation 
of the loop oscillations in time (at $\sim$19:40 UT).

\section{Summary and Discussion}
The SDO/AIA observations of transverse loop oscillations analyzed have revealed 
several interesting new features. The loop consists of multithermal
strands, whose dynamical behaviors are temperature-dependent. In the 171 \AA\ band, two strands 
show in-phase oscillations with growing amplitudes and a separating drift. 
Their displacement oscillations are associated with the intensity variations. In the 193 \AA\ band, 
two close strands show the oscillations with no clear amplitude change and a quarter-period phase 
delay has developed between them after a time of about two periods. 
The oscillations of these strands have very similar periods.

The flare-excited transverse loop oscillations observed by TRACE have been interpreted 
as eigenmodes (mainly the fundamental kink mode) \citep{nak99, asc02}. These oscillations 
typically show the strong damping, which has been suggested due to resonant absorption or 
wave leakage \citep[see reviews by][]{rob00, rud09}. The unusual growing oscillations 
reported here suggest that they may be forced kink oscillations with continuous energy 
input at a rate faster than the damping. The STEREO-A observations suggest 
that continuous interaction from the erupted flux rope in a CME may play the role of the
external driver. A theoretical study by \citet{bal08} showed that a harmonic driver
typically excites a mixture of standing kink modes harmonics (with both the driver's 
and the natural periods). The oscillations analyzed here show mostly a single frequency. 
This feature does not agree with the harmonic driver, and suggests the  
excitation by a continuous non-harmonic driver. However, we notice that the timing of the 
amplitude-increasing oscillations is coincident with both the CME dimming formation near 
the loop footpoint and the emergence of the surge. Although the flux rope eruption as the 
driver of the oscillations is the preferred interpretation as discussed in the last section, 
the possibility that the surge also could play a role in reinforcing the oscillations cannot 
be entirely excluded. Recently, quasi-periodic fast mode magnetosonic waves with a propagation 
speed of more than 2000 km~s$^{-1}$ and a total duration about 30 minutes were discovered 
by SDO/AIA \citep{liu11, ofm11}. Whether such waves were produced by the surge and could 
have driven the amplification of oscillations needs further investigations observationally 
and theoretically.

Some studies have suggested that the loop cooling can strongly affect the kink oscillations
\citep{asc08, mor09}. Recently, an undamped kink oscillation event was 
observed by SDO/AIA \citep{asc11}, and the lack of damping has been attributed to the cooling 
which can amplify the oscillation \citep{rud11a, rud11b, rud11c}. Note that for an unrealistic
assumption, \citet{mor09} obtained that cooling causes the damping of kink oscillations 
\citep{rud11a}.  Our discussion below based on 
the models developed by \citet{rud11a} suggests that the growing oscillations reported 
here cannot be explained by changes of the loop temperature with time.
For the sake of present discussion, we assume that the decrease of the loop intensity in the 
171 \AA\ band is due to the cooling (although our analysis above suggests heating 
of the loop strands). Considering the role of wave damping,
the measured amplitude growth time ($\tau_g$) should be the upper limit of the
amplification time ($t_{amp}$) due to the cooling. From the measured loop height ($h$=77 Mm), 
we obtain the parameter, $\kappa$=$h/H_0\approx2$, where $H_0$ is the atmospheric scale height
for the initial loop plasma temperature of about 1 MK. For the loop model with $\kappa$=2 and
$\chi\approx$0.1 (the ratio of the loop external and internal plasma densities), we obtain
$t_{amp}\approx 4t_{cool}$ from the dependence of the oscillation amplitude
on time for the model of stratified loop with constant temperature of external plasma
\citep[Equation~(38) and the corrected Figure 7 in][]{rud11a, rud11b}, where 
$t_{cool}$ is the loop cooling time. Since $t_{amp}<\tau_g$, we obtain $t_{cool}/P<$1.4 and 0.8 
for the upper and lower threads, respectively. This means that the cooling must be very
fast with the characteristic cooling time less than the oscillation period. For such a rapid
cooling the oscillating loop should become completely invisible in the 171 \AA\ band
after two periods, and the oscillation should show a dramatic ($>50\%$) decrease in period 
over the lifetime \citep{mor09, rud11a}. However, neither theoretical predictions are 
consistent with present observations. For the same reason the undamped oscillations
observed in the 193 \AA\ band are also impossible to interpret by the cooling effect 
when the typical finite damping rate by resonant absorption is considered \citep{rud11c}.
This disagreement supports our suggestion that the wave energy in the loop is supplied 
continuously during the oscillations in our case, in contrast with the initial impulsive 
excitation suggested by the typical damping scenario of resonant absorption.

The above discussions are based on the properties of monolithic tube models, 
whereas in our case the oscillating loop consists of multithermal
strands, thus the interpretation of their dynamic behaviors may need to
consider the properties of coupled multi-stranded loop models \citep[e.g.][]{ofm09, lun09}. 
Our observations show kink oscillations of several loop strands with similar frequencies and
phase shift of in-phase or quarter-period, providing the evidence
for a collective kink mode \citep{lun09}. 
\citet{lun08} showed the simultaneous excitation of several collective
normal modes can lead to a $\pi$/2 phase difference between two neighboring loops and 
the beating of the system in some cases. The setup of a quarter-period phase shift
between two 193 \AA\ strands may belong to such a case.  We notice 
that after the coupling the two 193 \AA\ strands oscillate
with a period slightly shorter (by $\sim$20\%) than before (assumed to be the kink-mode
period of the individual loops). This feature also agrees with the model-prediction 
\citep[see Equation~(9) in][]{lun08} when the two loops are very close 
(with the similar separation as observed). 
It is unclear whether the amplitude-growing oscillations of the 171 \AA\ strands are due to 
the particular combination of collective normal modes excited in certain condition as no
beating behavior was found as predicted \citep{lun08, lun10}. 
In addition, our observations show the evident temperature dependence 
of multistanded loop oscillations, which were not studied in previous models.

The association of loop displacement oscillations with intensity and loop width variations
is found in this study. However, the positive correlation between the loop 
width and intensity variations suggests that the loop width variations may be 
observational artifacts due to the line-of-sight intensity variations.
Assuming the mass conservation in the loop, an anti-correlation between
them (ie., $\Delta{I}/I\sim-3\Delta{d}/d$, where $I$ is the loop intensity, and
$d$ the loop diameter) is predicted theoretically \citep{asc11}. 
The associated intensity oscillations could be due to variations in the line-of-sight 
column depth of the oscillating loop as suggested in some previous studies
\citep{coo03, ver09, ver10, whi12}, but this conjecture needs
a forward modeling to confirm.

\acknowledgments
The authors are grateful to Drs. Jaume Terradas and Manuel Luna for their valuable 
comments. The work of TW was supported by NASA grants NNX08AE44G, NNX10AN10G,
and NNX12AB34G. LO acknowledges to support from NASA grants NNX09AG10G, NNX10AN10G, 
and NNX12AB34G.

\end{document}